\documentclass[12pt]{article}
\usepackage{amssymb}
\usepackage{amsfonts}
\usepackage{graphicx}
\begin{document}

\title{{\LARGE APPLICATION OF CHAOS DEGREE TO SOME DYNAMICAL\ SYSTEMS}}
\author{$\dagger $Kei Inoue, $\dagger $Masanori Ohya and $\ddagger $Keiko Sa
to \\
$\dagger $Department of Information Sciences,\\
Science University of Tokyo,\\
Noda City, Chiba 278-8510, Japan.\\
$\ddagger $Department of Control and Computer Engineering,\\
Numazu College of Technology,\\
Numazu City, Shizuoka 410-8501, Japan}
\date{}
\maketitle

\begin{abstract}
Chaos degree defined through two complexities in information dynamics is
applied to some deterministic dynamical models. It is shown that this degree
well describes the chaostic feature of the models.
\end{abstract}

\section{Introduction}

There exist several approaches in the study of chaotic behavior of dynamical
systems using the concepts such as entropy, complexity, chaos, fractal,
stochasticity \cite{Aka,Ali,Ben,Dev,Has,Tod}. In 1991, one of the authors
proposed Information Dynamics (ID for short) \cite{O1,O2} to treat such
chaotic behavior of systems altogether. ID is a synthesis of the dynamics of
state change and the complexity of systems, and it is applied to several
different fields such as quantum physics, fractal theory, quantum
information and genetics\cite{IKO}.

A quantity measuring chaos in dynamical systems was defined by means of two
complexities in ID, and it is called chaos degree. In particular, among
several chaos degrees, an entropic chaos degree was introduced in \cite
{O3,KO}, and is applied to logistic map to study its chaotic behavior. This
chaos degree has several merits compared with usual measures of chaos such
as Lyapunov exponent.

In Section2, we review the complexity in ID and the chaos degree (CD for
short). In Section 3, we remind the entropic chaos degree and the Lyapunov
exponent (LE for short). In Section 4, the algorithm computing the entropic
chaos degree is shown. In Section 5, we compute CD and LE for Bernoulli
shift, Baker's transformation and Tinkerbell's map, then we discuss the
merits of the entropic chaos degree.

\section{Complexity of Information Dynamics and Chaos Degree}

Information dynamics provides a frame to study the state change and the
complexity associated with a dynamical system. We briefly explain the
concept of the complexity of ID in a bit simplified version (see\cite{O2}).

Let $(\mathcal{A},{\frak S},\alpha (G))$ be an input
(or initial) system and $%
(\overline{\mathcal{A}},\overline{{\frak
S}},\overline{\alpha }(\overline{G} ))$ be an output
(or final) system. Here $\mathcal{A}$ is the set of
all objects to be observed and ${\frak S}$ is
the set of all means for measurement of
$\mathcal{A}$, $\alpha (G)$ is a certain evolution
of system. Often we have
$\mathcal{A}=\overline{\mathcal{A}}$,
${\frak S}=\overline{{\frak S}}$, $\alpha
=\overline{\alpha }$.

\smallskip For instance, when $\mathcal{A}$ is the set $M(\Omega )$ of all
measurable functions on a measurable space $(\Omega ,\mathcal{F})$ and $%
{\frak S}(\mathcal{A})$ is the set $P(\Omega )$ of all
probability measures on $\Omega $, we have usual
probability theory, by which the classical dynamical
system is described. When $\mathcal{A}$ is the set
$B(\mathcal{H})$ of all bounded linear operators on a
Hilbert space $\mathcal{H}$ and ${\frak S}(\mathcal{A})$
is the set ${\frak S}(\mathcal{H})$ of all density
operators on $\mathcal{H}$, we have a quantum dynamical
system.

Once an input and an output systems are set, the situation of the input
system is described by a state, an element of ${\frak S}$
$,$ and the change of the state is expressed by a mapping
from ${\frak S}$ to $\overline{{\frak S}}$,
called a channel. The concept of channel is fundamental
both in physics and mathematics \cite{IKO}. Moreover,
there exist two complexities in ID, which are
axiomatically given as follows:

Let $(\mathcal{A}_{t},{\frak S}_{t},\alpha ^{t}(G^{t}))$ be
the total system of $(\mathcal{A},{\frak S},\alpha )$
and
$(\overline{\mathcal{A}},\overline{{\frak S}},\overline{\alpha
}),$ and let $C\left( \varphi \right) $ be the complexity of
a state $\varphi $ and $T\left(
\varphi ;\Lambda ^{*}\right) $ be the transmitted
complexity associated with the state change$\;\varphi \to
\Lambda ^{*}\varphi .$ These complexities $C$ and $T$ are the quantities
satisfying the following conditions:

\begin{enumerate}
\item[(i)]  For any $\varphi \in {\frak S}$,
\[
C(\varphi )\ge 0,\ T(\varphi ;\Lambda ^{*})\ge 0.
\]

\item[(ii)]  For any orthogonal bijection
$j:ex{\frak S}\mathcal{\rightarrow } ex{\frak S}$ ( the set of
all extreme points in ${\frak S}$ ),
\[
C(j(\varphi ))=C(\varphi ),
\]
\[
T(j(\varphi );\Lambda ^{*})=T(\varphi ;\Lambda ^{*}).
\]

\item[(iii)]  For $\Phi \equiv \varphi \otimes \psi \in
{\frak S}_{t}$,
\[
C(\Phi )=C(\varphi )+C(\psi ).
\]

\item[(iv)]  For any state $\varphi $ and a channel $\Lambda ^{*},$%
\[
0\le T(\varphi ;\Lambda ^{*})\le C(\varphi ).
\]

\item[(v)]  For the identity map ``id'' from ${\frak S}$ to
${\ }{\frak S}$.
\[
T(\varphi ;id)=C(\varphi ).
\]
\end{enumerate}

When a state $\varphi $ changes to the state $\Lambda ^{*}\varphi $, a $%
\emph{chaos}$\emph{\ degree} (CD) \cite{O2} w.r.t. $\varphi $ and $\Lambda
^{*}$ is given by

\[
D\left( \varphi ;\Lambda ^{*}\right) =C\left( \Lambda ^{*}\varphi \right)
-T\left( \varphi ;\Lambda ^{*}\right) .
\]
Using the above CD, we observe chaos of a dynamical system as
\begin{eqnarray*}
CD &>&0\Longleftrightarrow \mbox{chaotic.} \\
CD &=&0\Longleftrightarrow \mbox{stable.}
\end{eqnarray*}

\section{Entropic Chaos Degree and Lyapunov Exponent}

Chaos degree in ID was applied to a smooth map on $R$ and it is shown that
this degree enables to describe the chaotic aspects of a logistic map as well%
\cite{KO,O3}. Here we briefly review the chaos degree defined through
classical entropies. For an input state described by a probability
distribution $p=\left( p_{i}\right) $ and the joint distribution $r=\left(
r_{i,j}\right) $ between $p$ and the output state $\bar{p}=\Lambda
^{*}p=\left( \overline{p}_{i}\right) $ through a channel $\Lambda ^{*}$, the
Shannon entropy $S\left( p\right) =-\sum_{i}p_{i}\log p_{i}$ and the mutual
entropy $I\left( p;\Lambda ^{*}\right) =\sum_{i,j}r_{i,j}\log \frac{r_{i,j}}{
p_{i}\bar{p}_{j}}$ satisfy all conditions of the complexities $C\ $and $T,$
then the entropic chaos degree is defined by

\begin{eqnarray*}
D\left( p;\Lambda ^{*}\right) &\equiv &C\left( \Lambda ^{*}p\right) -T\left(
p;\Lambda ^{*}\right) \\
&\equiv &S\left( \bar{p}\right) -I\left( p;\Lambda ^{*}\right) ,
\end{eqnarray*}
This entropic chaos degree is nothing but the conditional entropy of
aposteriori state w.r.t. the channel $\Lambda ^{*}$. The characteristic
point of the entropic chaos degree is easy to get the probability
distribution of the orbit for a deterministic dynamics, which is discussed
in the next section.

Lyapunov exponent (LE) is used to study chaotic behavior of a deterministic
dynamics. The Lyapunov exponent $\lambda \left( f\right) $ for a smooth map $%
f$ on $\bf{R}$ is defined by \cite{ASY}:

\[
\lambda \left( f\right) =\lim_{n\rightarrow \infty }\lambda _{n}\left( f\
\right) \,\, \lambda _{n}\left( f\right) =\frac{1}{n}\sum_{k=1}^{n}\log
\left| \frac{df}{dx}\left( x^{\left( k\right) }\right) \right| ,
\]
where $x^{\left( n\right) }=f\left( x^{\left( n-1\right) }\right) $ for any $%
n\in \bf{N.}$

For a smooth map $f=\left( f_{1},\cdots ,f_{m}\right) $ on $\bf{R}^{m},$
the vector version of LE is defined as follows: Let $\ x^{0}$ be an initial
point of $\bf{R}^{m}$ and $x^{\left( n\right) }=f\left( x^{\left(
n-1\right) }\right) $ for any $n\in \bf{N.}$ After $n$ times iterations
of $f$ to $x^{0,}$ the Jacobi matrix $J_{n}\left( x^{0}\right) $ of $%
x^{\left( n\right) }=\left( x_{1}^{\left( n\right) },\cdots ,x_{m}^{\left(
n\right) }\right) $w.r.t. $x^{\left( 0\right) }$is

\[
J_{n}\left( x^{(0)}\right) =Df^{n}\left( x^{(0)}\right) =\left[
\begin{array}{lll}
\frac{\partial f_{1}^{\left( n\right) }}{\partial x_{1}}\left( x^{\left(
0\right) }\right) & \cdots & \frac{\partial f_{1}^{\left( n\right) }}{%
\partial x_{m}}\left( x^{\left( 0\right) }\right) \\
\cdots & \cdots & \cdots \\
\frac{\partial f_{m}^{\left( n\right) }}{\partial x_{1}}\left( x^{\left(
0\right) }\right) & \cdots & \frac{\partial f_{m}^{\left( n\right) }}{%
\partial x_{m}}\left( x^{\left( 0\right) }\right)
\end{array}
\right] ,
\]
Then the Lyapunov exponent $\lambda \left( f\right) $ of $x^{\left( 0\right)
}$ is defined by

\[
\lambda \left( f\right) =\log \bar{\mu}_{1},\bar{\mu}_{k}=\lim_{n\rightarrow
\infty }\left( \mu _{k}^{n}\right) ^{\frac{1}{n}}(k=1,2,\cdot \cdot \cdot
,m).
\]
Here $\mu _{k}^{n}$ is the $k$th largest square root of the eigenvalues of
the matrix $J_{n}\left( x^{\left( 0\right) }\right) J_{n}\left( x^{\left(
0\right) }\right) ^{T}$.

An orbit of the dynamical system described by $f$ is said to be chaotic when
the exponent $\lambda \left( f\right) $ is positive, and to be stable when
the exponent is negative. The positive exponent means that the orbit is very
sensitive to the initial value, so that it describes a chaotic behavior.
Lyapunov exponent is difficult to compute for some models (e.g., Tinkerbell
map) and its negative value is not clearly explained.

\section{Algorithm for Computation of Entropic \\ Chaos Degree}

It is proved \cite{Mis} that if a piecewise monotone mapping $f$ from $%
\left[ a,b\right] ^{m}\ $to $\left[ a,b\right] ^{m}$ has non-positive
Schwarzian derivatives and does not have a stable and periodic orbit, then
there exists an ergodic measure $\mu $ on the Borel set
${\frak S}$ of $%
\left[ a,b\right] ^{m}$, absolutely continuous w.r.t. the Lebesgue measure.

Take a finite partition $\left\{ A_{k}\right\} $ of $I=\left[ a,b\right]
^{m} $ such as

\[
I=\bigcup_{k}A_{k}\quad \left( A_{i}\bigcap A_{j}=\emptyset ,i\neq j\right)
.
\]

\noindent Let $\left| S\right| $ be the number of the elements in a set $S$.
Suppose that $n$ is sufficiently large natural number and $m$ is a fixed
natural number. Let $p^{\left( n\right) }\equiv \left( p_{i}^{\left(
n\right) }\right) $ be the probability distribution of the orbit up to $n$
-th step, that is, how many $x^{(k)}\left( k=m+1,\cdots ,m+n\right) $ are in $%
A_{i}$:

\[
p_{i}^{\left( n\right) }\equiv \frac{\left| \left\{ k\in {\bf N}
;x^{\left( k\right) }\in A_{i},m<k\leq m+n\right\} \right| }{n}
\]

It is shown that the $n\rightarrow \infty $ limit of $p_{i}^{\left( n\right)
}$ exists and equals to $\mu \left( A_{i}\right) $. The channel $\Lambda
^{*} $ is a map given by $p^{\left( n+1\right) }=\Lambda ^{*}p^{\left(
n\right) }$. Further, the joint distribution $r^{\left( n,n+1\right)
}=\left( r_{i,j}^{\left( n,n+1\right) }\right) $for a sufficient large $n$
is approximated as

\[
r_{i,j}^{\left( n,n+1\right) }=\frac{\left| \left\{ k\in {\bf N};\left(
x^{\left( k\right) },f\left( x^{\left( k\right) }\right) \right) \in
A_{i}\times A_{j},m<k\leq m+n\right\} \right| }{n}.
\]

Then the entropic chaos degree is computed as

\begin{eqnarray*}
D\left( p^{\left( n\right) };\Lambda ^{*}\right) &=&S\left( p^{\left(
n+1\right) }\right) -I\left( p^{\left( n\right) };\Lambda ^{*}\right) \\
&=&-\sum_{i}p_{i}^{(n+1)}\log p_{i}^{(n+1)}-\sum_{i,j}r_{i,j}^{\left(
n,n+1\right) }\log \frac{r_{i,j}^{\left( n,n+1\right) }}{p_{i}^{\left(
n\right) }p_{j}^{(n+1)}}
\end{eqnarray*}

\section{Entropic Chaos Degree for Some Determi-nistic Dynamical
Models}

In this section, we study the chaotic behavior of several well-known
deterministic maps by the entropic chaos degree.

\noindent
\subsection*{$\langle$1$\rangle$ Bernoulli shift}

Let $f$ a map from $\left[ 0,1\right] \ $to itself such that

\begin{equation}
\renewcommand{\theequation}{5.1}
f\left( x_{n}\right) =\left\{
\begin{array}{l}
2ax^{\left( n\right) } \\
a\left( 2x^{\left( n\right) }-1\right)
\end{array}
\begin{array}{l}
\left( 0\leq x^{\left( n\right) }\leq 0.5\right) \\
\left( 0.5<x^{\left( n\right) }\leq 1\right)
\end{array}
,\right.
\end{equation}
where $x^{\left( n\right) }\in \left[ 0,1\right] $ and $0\leq a\leq 1$.

Let us compute the Lyapunov exponent and the entropic chaos degree (ECD for
short) for the above Bernoulli shift $f$ .

The orbit of the equation (5.1) is shown in Fig. 1:

%\begin{figure}[H]
\begin{center}
\includegraphics[width=8cm,height=5cm]
{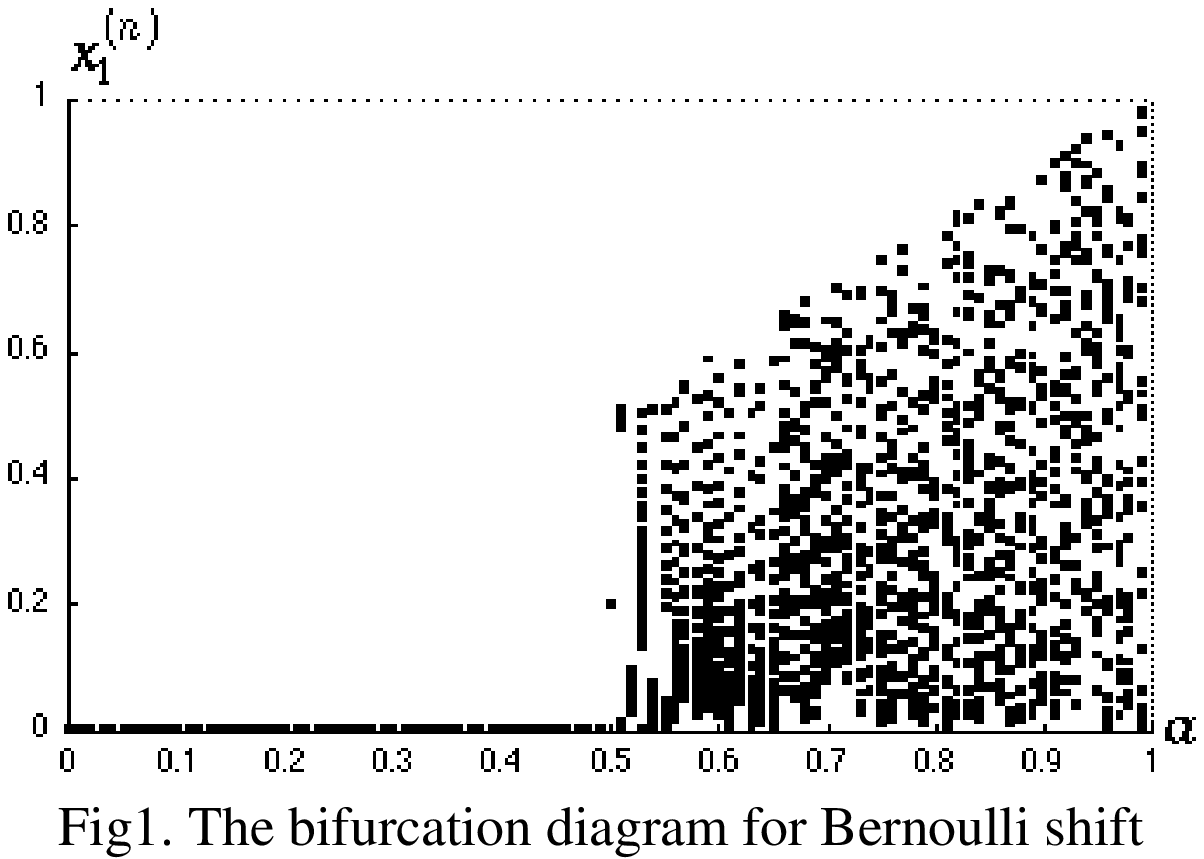}
%\caption{The bifurcation diagram for Bernoulli shift}
\end{center}
%\end{figure}

The Lyapunov exponent $\lambda _{n}\left( f\right) $ is $\log 2a$ for the
Bernoulli shift (Fig. 2).
%\begin{figure}[H]
\begin{center}
\includegraphics[width=8cm,height=5cm]
{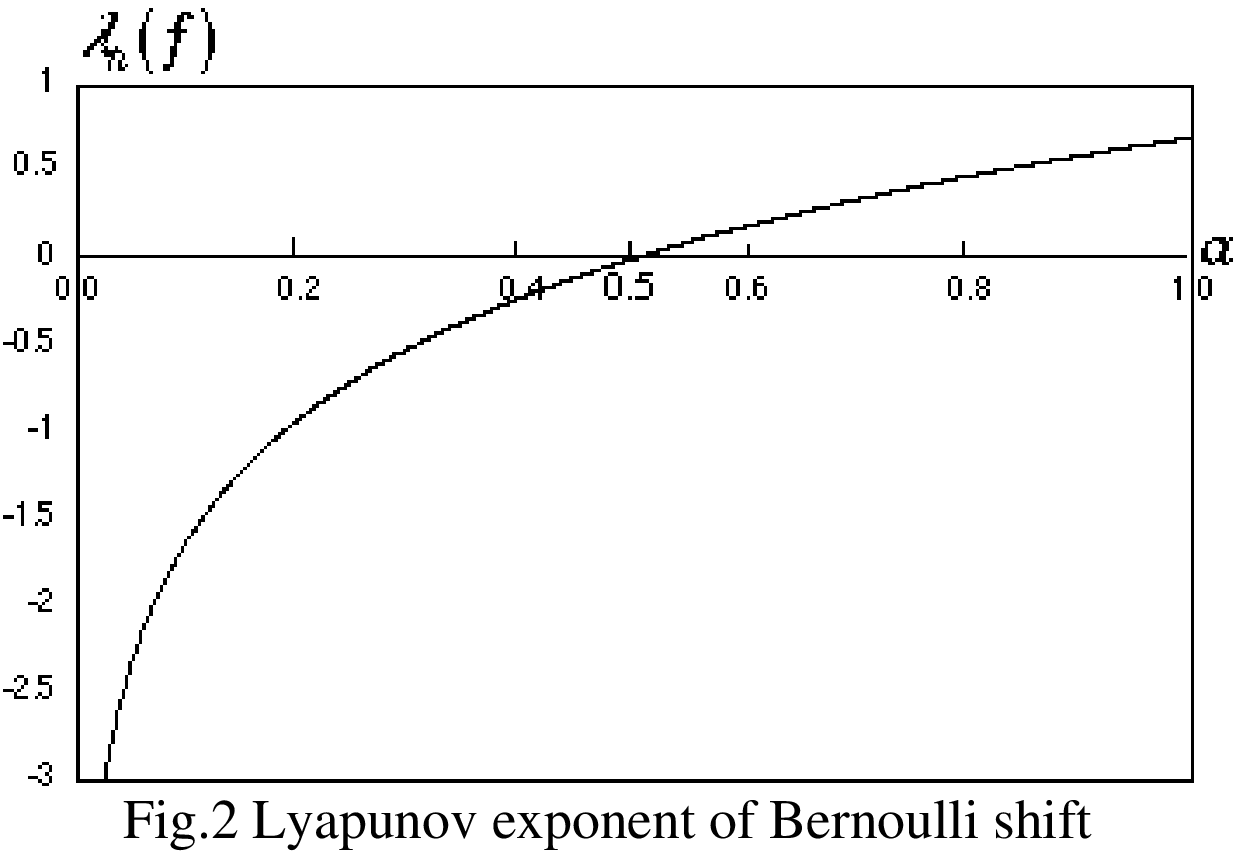}
%\caption{Lyapunov exponent of Bernoulli shift}
\end{center}
%\end{figure}

On the other hand, the entropic chaos degree of the Bernoulli shift is
shown in Fig. 3.
%\begin{figure}[H]
\begin{center}
\includegraphics[width=8cm,height=5cm]
{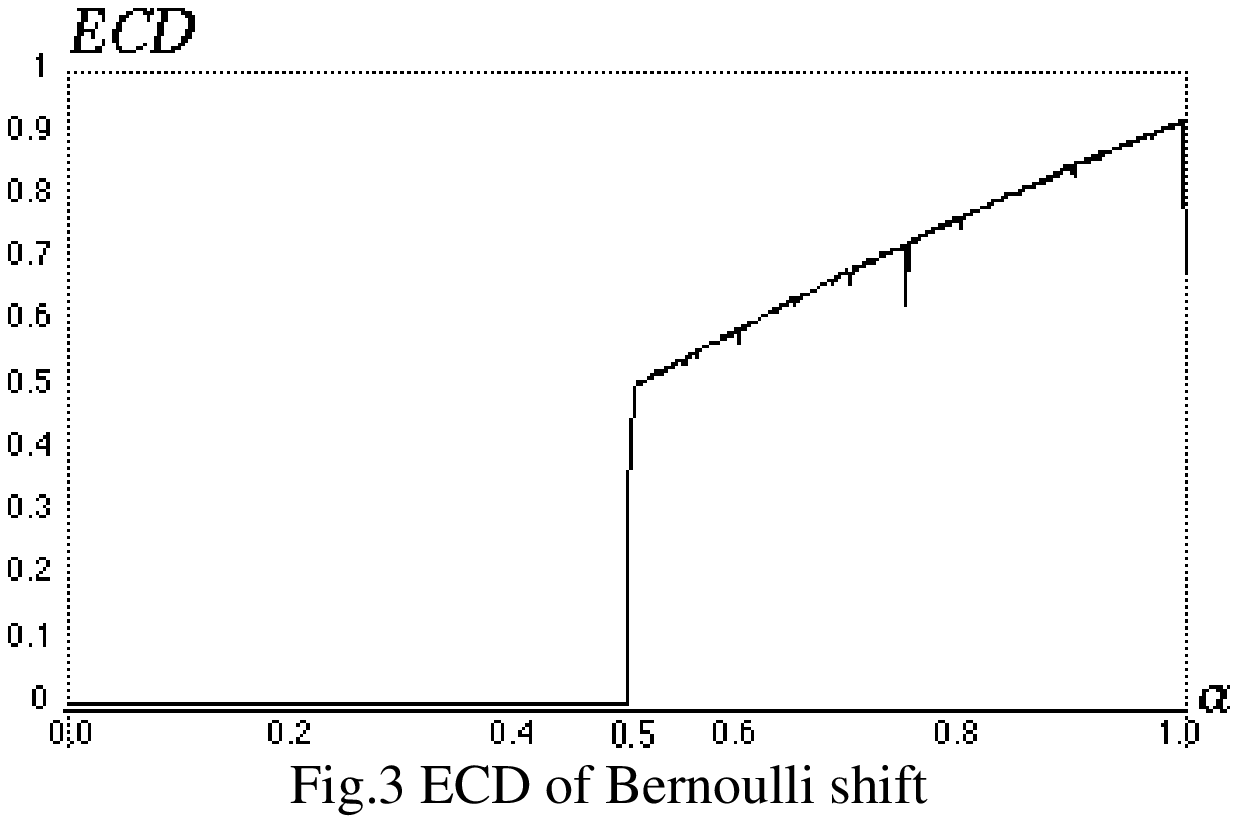}
%\caption{ECD of Bernoulli shift}
\end{center}
%\end{figure}

Here we took $740$ different $a$'s between $0$ and $1$ with

\begin{eqnarray*}
A_{i} &=&\left[ \frac{i}{2000},\frac{i+1}{2000}\right] \quad \left(
i=0,\cdots ,1999\right) \\
n &=&100000.
\end{eqnarray*}

\noindent
\subsection*{$\langle$2$\rangle$ Baker's transformation }

We apply the chaos degree to a smooth map on $R^{2}$. Let us compute the
Lyapunov exponent and the ECD for the following Baker's transformation $%
f_{a} $ :

\begin{eqnarray*}
f_{a}\left( x^{\left( n\right) }\right)& = &f_{a}\left( x_{1}^{\left( n\right)
},x_{2}^{\left( n\right) }\right)\\
& = &\left\{
\begin{array}{l}
\left( 2ax_{1}^{\left( n\right) }\,\, \frac{1}{2}ax_{2}^{\left( n\right)
}\right) \\
\left( a\left( 2x_{1}^{\left( n\right) }-1\right) ,\frac{1}{2}a\left(
x_{2}^{\left( n\right) }+1\right) \right)
\end{array}
\right.
\begin{array}{l}
\left( 0\leq x_{1}^{\left( n\right) }\leq 0.5\right) \\
\left( 0.5<x_{1}^{\left( n\right) }\leq 1\right)
\end{array},
\end{eqnarray*}

\noindent where $\left( x_{1}^{\left( n\right) },x_{2}^{\left( n\right)
}\right) \in \left[ 0,1\right] \times \left[ 0,1\right] $ and $0\leq a\leq 1$
.\newpage

The orbit for each $a$ is shown in Fig. 4. $\sim$9.

%\begin{figure}[H]
\begin{center}
\includegraphics[width=8cm,height=5cm]
{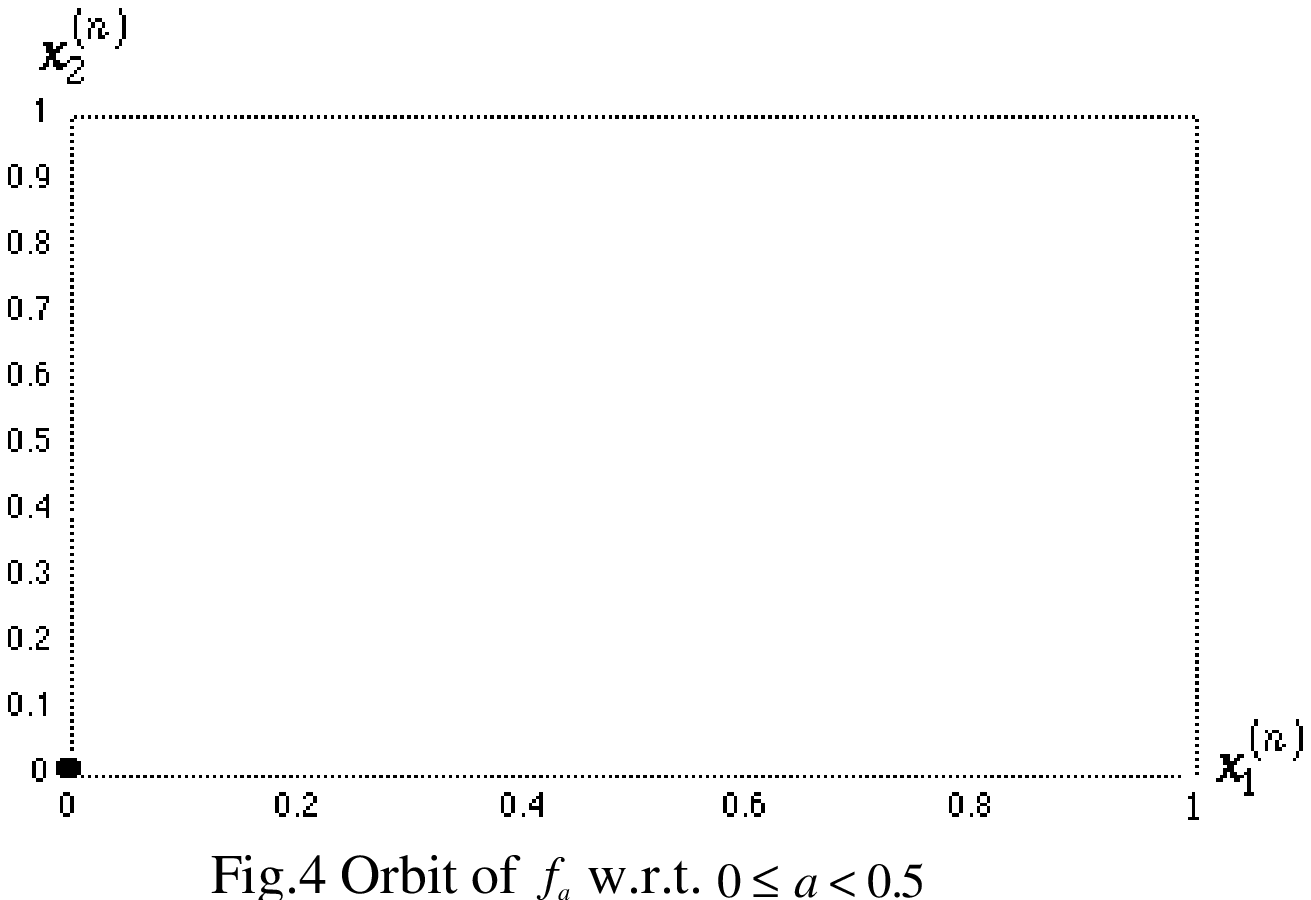}
%\caption{Orbit of $f_{a}$ w.r.t. $0\leq a<0.5$}
\end{center}
%\end{figure}

%\begin{figure}[H]
\begin{center}
\includegraphics[width=8cm,height=5cm]
{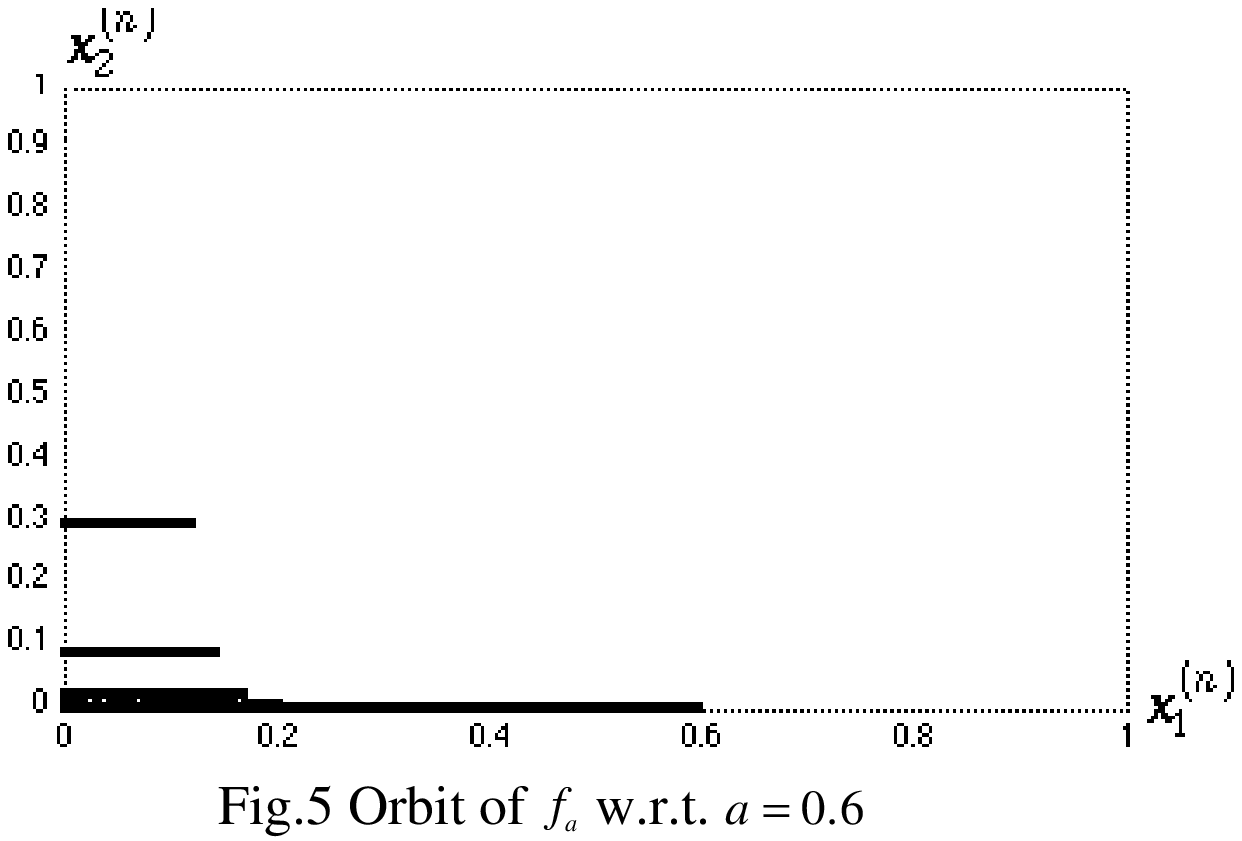}
%\caption{Orbit of $f_{a}$ w.r.t. $a=0.6$}
\end{center}
%\end{figure}

%\begin{figure}[H]
\begin{center}
\includegraphics[width=8cm,height=5cm]
{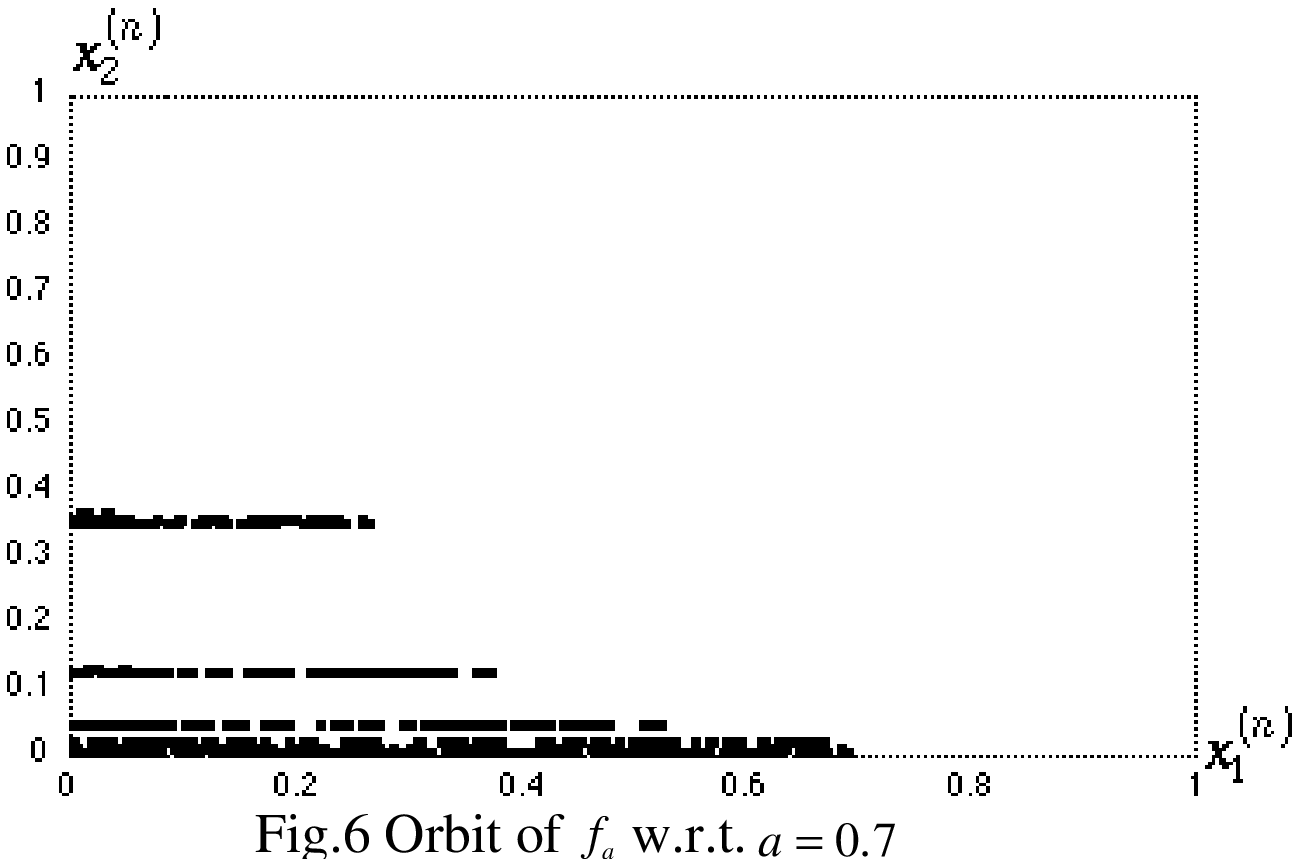}
%\caption{Orbit of $f_{a}$ w.r.t. $a=0.7$}
\end{center}
%\end{figure}

%\begin{figure}[H]
\begin{center}
\includegraphics[width=8cm,height=5cm]
{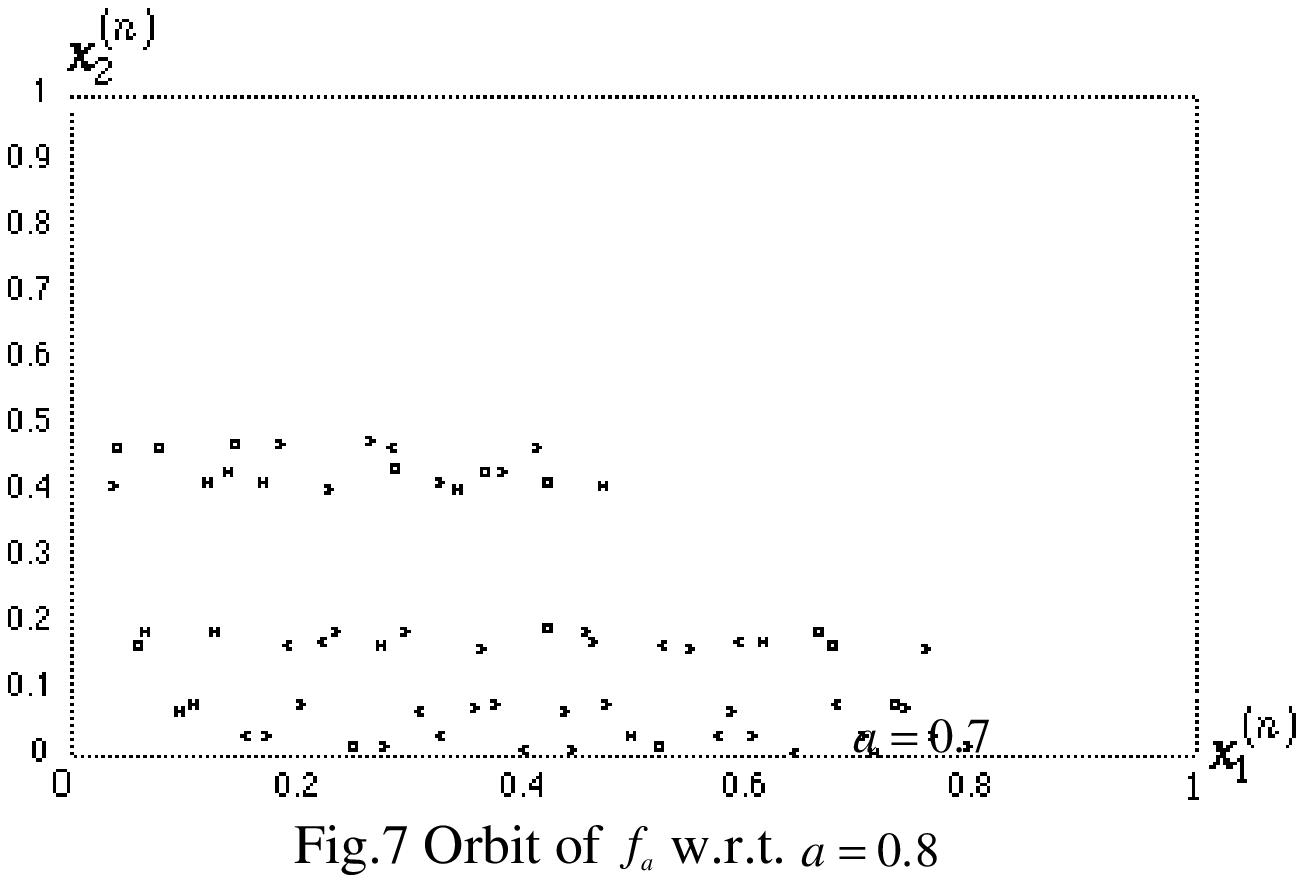}
%\caption{Orbit of $f_{a}$ w.r.t. $a=0.8$}
\end{center}
%\end{figure}

%\begin{figure}[H]
\begin{center}
\includegraphics[width=8cm,height=5cm]
{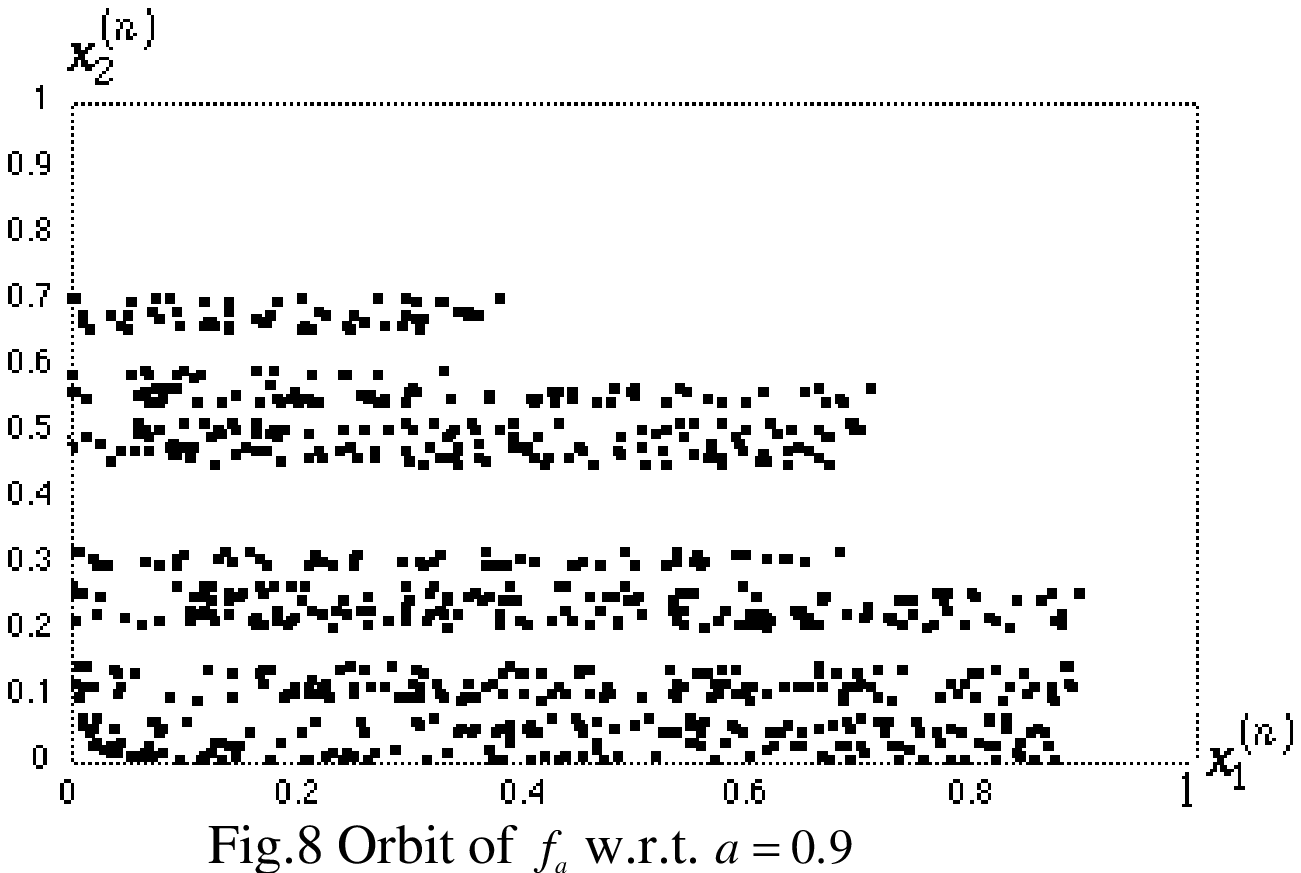}
%\caption{Orbit of $f_{a}$ w.r.t. $a=0.9$}
\end{center}
%\end{figure}

%\begin{figure}[H]
\begin{center}
\includegraphics[width=8cm,height=5cm]
{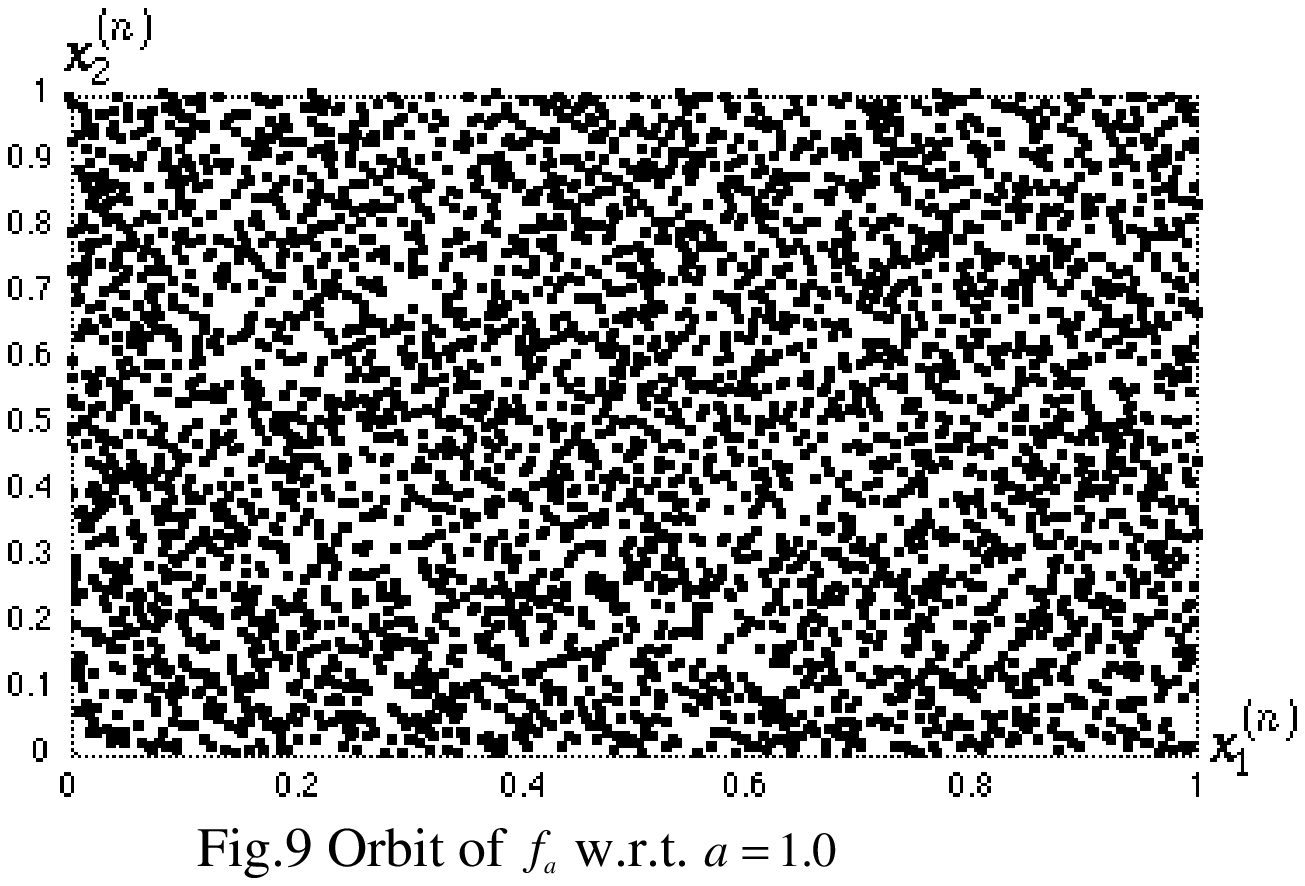}
%\caption{Orbit of $f_{a}$ w.r.t. $a=1.0$}
\end{center}
%\end{figure}

These figures show that the larger $a$ is, the more complicated the orbit
is. The maximum Lyapunov exponent $\lambda _{n}^{1}\left( f\right) $ is
$\log 2a$ for Baker's transformation (Fig.10).

%\begin{figure}[H]
\begin{center}
\includegraphics[width=8cm,height=5cm]
{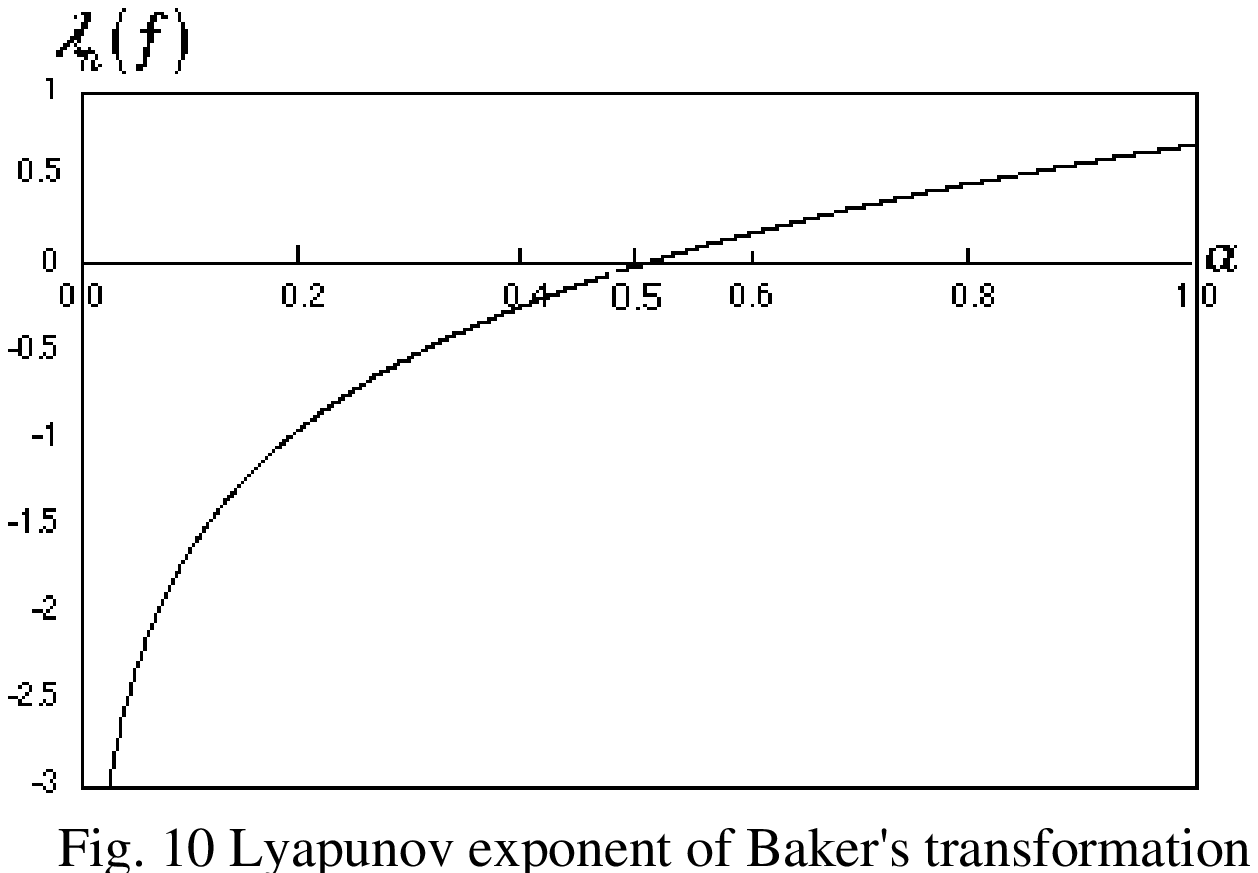}
%\caption{Lyapunov exponent of Baker's transformation}
\end{center}
%\end{figure}

On the other hand, the ECD of Baker's transformation is shown in
Fig.11.

%\begin{figure}[H]
\begin{center}
\includegraphics[width=8cm,height=5cm]
{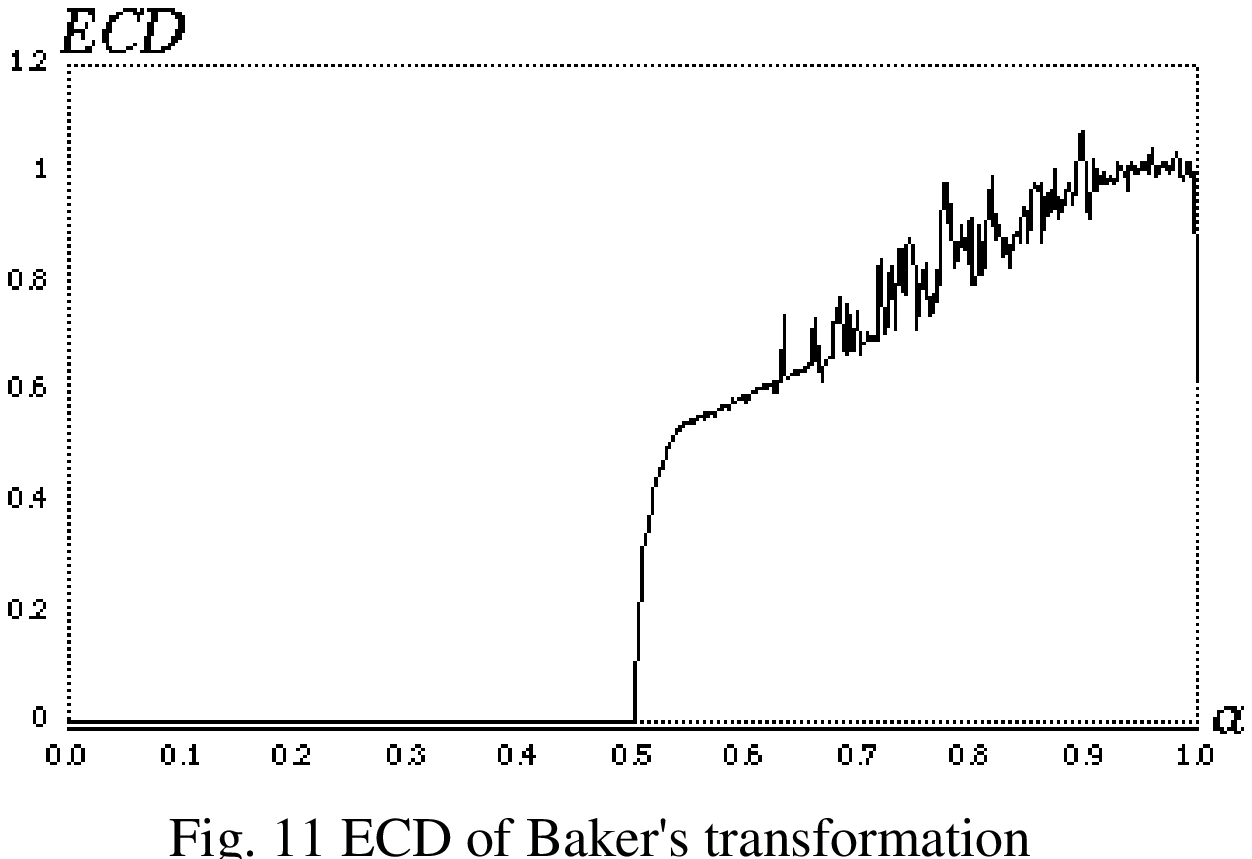}
%\caption{ECD of Baker's transformation}
\end{center}
%\end{figure}

Here we took $740$ different $a$'s between $0$ and $1$ with

\begin{eqnarray*}
A_{i,j} &=&\left[ \frac{i}{100},\frac{i+1}{100}\right] \times \left[ \frac{j%
}{100},\frac{j+1}{100}\right] \quad \left( i,j=0,\cdots ,99\right) \\
n &=&100000.
\end{eqnarray*}

\noindent
\subsection*{$\langle$3$\rangle$ Tinkerbell map}

\bigskip

Let us compute the CD for the following two type Tinkerbell maps
$f_{a}$ and $f_{b}$ on $I=\left[ -1.2,0.4\right] \times \left[
-0.7,0.3\right] $.

\begin{eqnarray*}
f_{a}\left( x^{\left( n\right) }\right)
&=&f_{a}\left( x_{1}^{\left( n\right) },x_{2}^{\left( n\right) }\right)\\
&=&\left( \left( x_{1}^{\left( n\right) }\right) ^{2}-\left( x_{2}^{\left(
n\right) }\right) ^{2}+ax_{1}^{\left( n\right) }+c_{2}x_{2}^{\left( n\right)
},2x_{1}^{\left( n\right) }x_{2}^{\left( n\right) }+c_{3}x_{1}^{\left(
n\right) }+c_{4}x_{2}^{\left( n\right) }\right) , \\
f_{b}\left( x^{\left( n\right) }\right)
&=&f_{b}\left( x_{1}^{\left( n\right) },x_{2}^{\left( n\right) }\right)\\
&=&\left( \left( x_{1}^{\left( n\right) }\right) ^{2}-\left( x_{2}^{\left(
n\right) }\right) ^{2}+c_{1}x_{1}^{\left( n\right) }+c_{2}x_{2}^{\left(
n\right) },2x_{1}^{\left( n\right) }x_{2}^{\left( n\right) }+bx_{1}^{\left(
n\right) }+c_{4}x_{2}^{\left( n\right) }\right) ,
\end{eqnarray*}

\noindent where $\left( x_{1}^{\left( n\right) },x_{2}^{\left( n\right)
}\right) \in I,-0.4\leq a\leq 0.9,1.9\leq b\leq 2.9,\left(
c_{1},c_{2},c_{3},c_{4}\right) =\left( -0.3,-0.6,2.0,0.5\right) ,$ and $%
\left( x_{1}^{\left( 0\right) },x_{2}^{\left( 0\right) }\right) =\left(
0.1,0.1\right) .$

Let us plot the points $\left( x_{1}^{\left( n\right) },x_{2}^{\left(
n\right) }\right) $ for $3000$ different $n$'s between 1001 and 4000.

In stable domain, the number of the points $\left( x_{1}^{\left( n\right)
},x_{2}^{\left( n\right) }\right) $ is finite because the point $\left(
x_{1}^{\left( n\right) },x_{2}^{\left( n\right) }\right) $ periodically
appears in time $n$. Fig.12 and Fig.13 are examples of the orbits of
$f_{a}$ and $f_{b}$ in a stable domain.

%\begin{figure}[H]
\begin{center}
\includegraphics[width=8cm,height=5cm]
{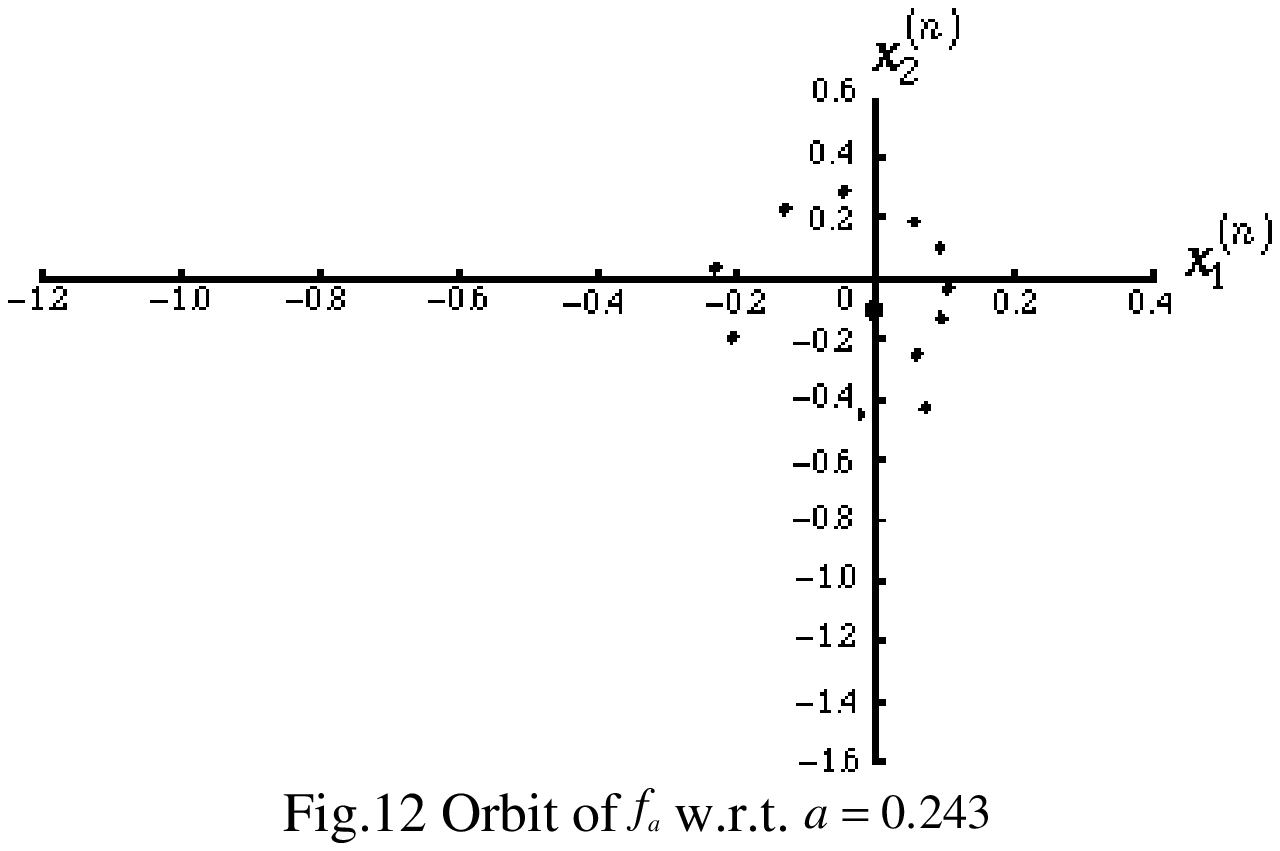}
%\caption{Orbit of $f_{a}$ w.r.t. $a=0.243$}
\end{center}
%\end{figure}

%\begin{figure}[H]
\begin{center}
\includegraphics[width=8cm,height=5cm]
{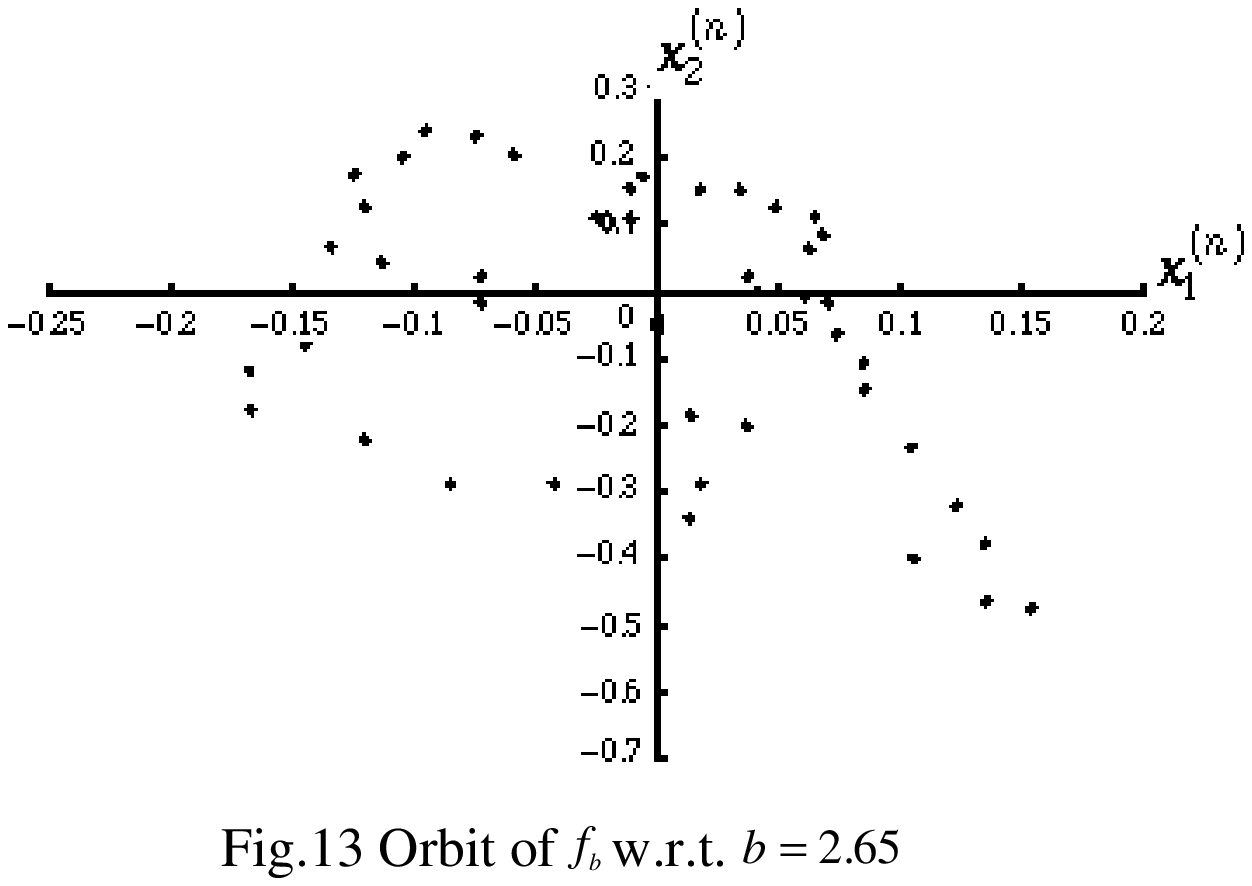}
%\caption{Orbit of $f_{b}$ w.r.t. $b=2.65$}
\end{center}
%\end{figure}

On the other hand, the point $\left( x_{1}^{\left( n\right) },x_{2}^{\left(
n\right) }\right) $ take random value in time $n$ in chaotic domain.
Fig.514 and Fig.15 are examples of the orbits of $f_{a}$ and $f_{b}$
in a chaotic domain.

%\begin{figure}[H]
\begin{center}
\includegraphics[width=8cm,height=5cm]
{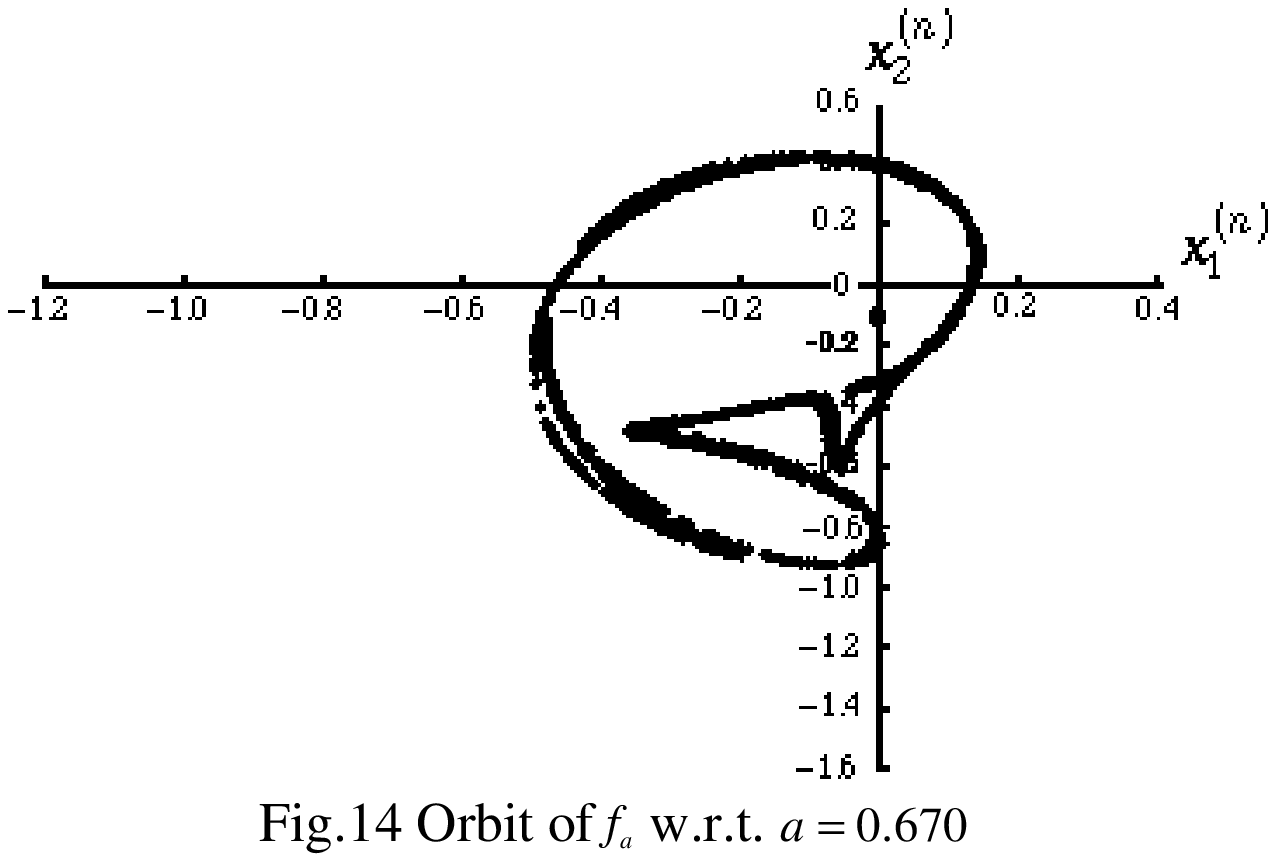}
%\caption{Orbit of $f_{a}$ w.r.t. $a=0.670$}
\end{center}
%\end{figure}

%\begin{figure}[H]
\begin{center}
\includegraphics[width=8cm,height=5cm]
{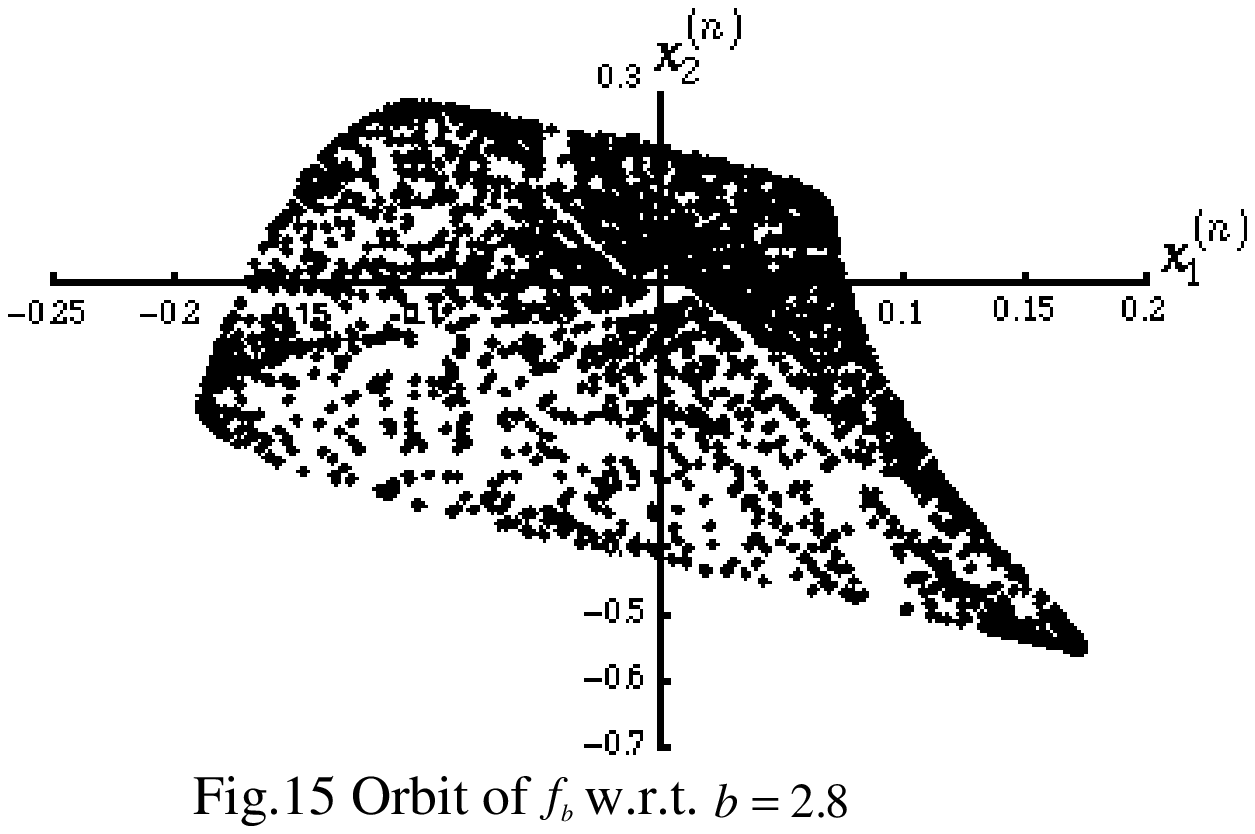}
%\caption{Orbit of $f_{b}$ w.r.t. $b=2.8$}
\end{center}
%\end{figure}

The ECD of Tinkerbell map $f_{a}$ and $f_{b}$ are shown in Fig. 16
and Fig. 17.

%\begin{figure}[H]
\begin{center}
\includegraphics[width=8cm,height=6cm]
{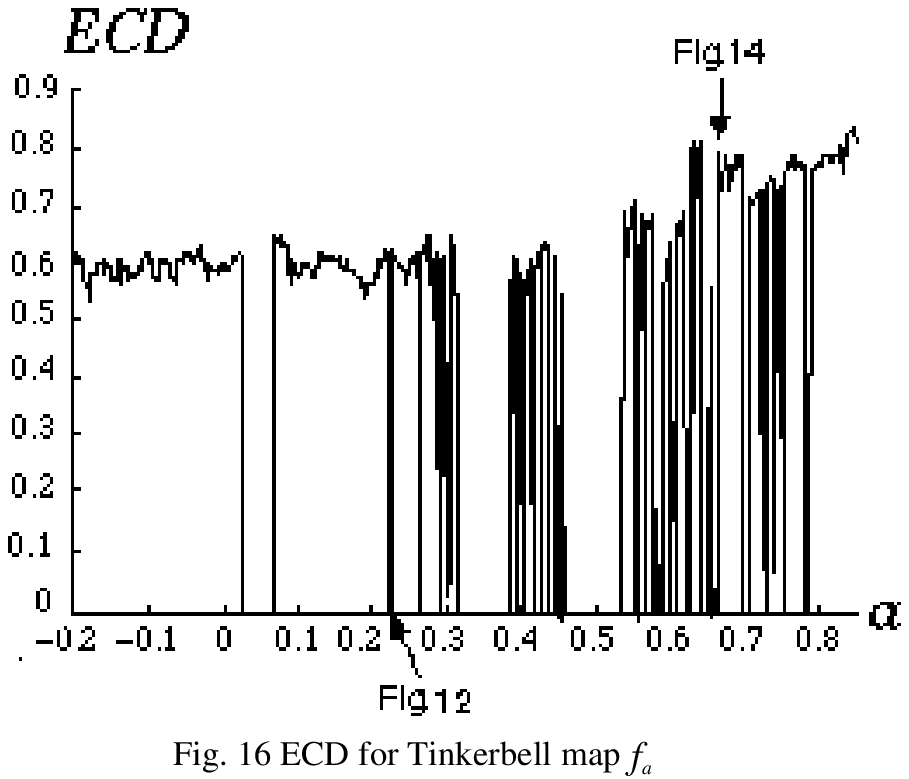}
%\caption{ECD for Tinkerbell map $f_{a}$}
\end{center}
%\end{figure}

%\begin{figure}[H]
\begin{center}
\includegraphics[width=8cm,height=6cm]
{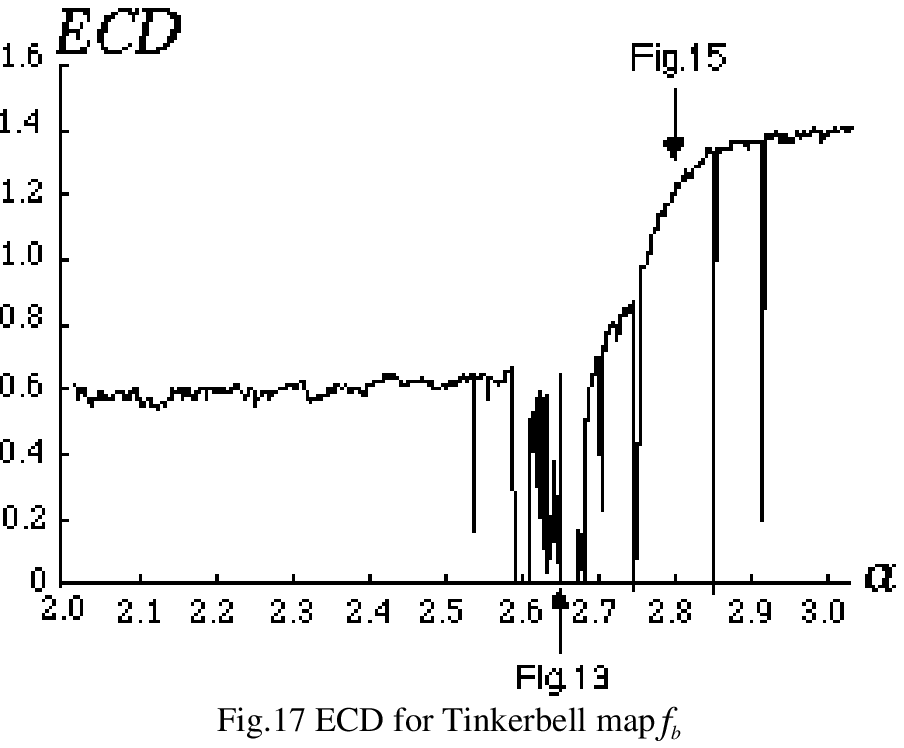}
%\caption{ECD for Tinkerbell map $f_{b}$}
\end{center}
%\end{figure}

Here we took $740$ different $a$'s between $-1.2$ and $0.9$ and $740$
different $b$'s between $1.9$ and $2.9$ with

\begin{eqnarray*}
A_{i,j} &=&\left[ \frac{i}{100},\frac{i+1}{100}\right] \times \left[ \frac{j%
}{100},\frac{j+1}{100}\right] \quad \\
&&\left( i=-120,-119,\cdots ,-1,0,1,\cdots ,38,39\right) \\
&&\left( j=-70,-69,\cdots ,-1,0,1,\cdots ,28,29\right) \\
n &=&100000.
\end{eqnarray*}

\section{Conclusion}

From our results, when the orbits for Bernoulli shift, Baker's
transformation or Tinkerbell map are chaotic, both Lyapunov exponent and
chaos degree are positive. However, our chaos degree can resolve some
inconvenient properties of the Lyapunov exponent in the following senses:

\begin{itemize}
\item[(1)]  Lyapunov exponent takes negative and sometimes $-\infty $. For
instance, although the exponent of Bernoulli shift and Baker's
transformation can not be defined for $a=0$, the ECD is always positive and
defined for any $a\geq 0$.

\item[(2)]  It is difficult to compute the Lyapunov exponent for the
Tinkerbell maps $f_{a}$ and $f_{b}$ because it is difficult to
compute $f_{a}^{n}$ and $f_{b}^{n}$ for large $n$. On the other
hand, the ECD of $f_{a}$ and $f_{b}$ are easily computed.

\item[(3)]  Generally, the algorithm for CD is much easier than that for
Lyapunov exponent.
\end{itemize}


\begin{thebibliography}{99}
\bibitem{Aka}  Akashi, S., The asymptotic behavior of $\varepsilon
$-entropy of a compact positive operator, J.Math.Anal.Appl., {\bf
153}, 250, 1990.

\bibitem{ASY} Alligood, K.T., Sauer, T.D., and  Yorke, J.A.,
\emph{Chaos-An Introduction to Dynamical Systems-}, Textbooks in
Mathematical Sciences, Springer, 1996)

\bibitem{Ali}  Alicki, R., Quantum geometry of noncommutative Bernoulli
shifts, Banach Center Publications, Mathematics Subject Classification
46L87,  1991.

\bibitem{Ben}  Bennatti, F.,  \emph{Deterministic Chaos in Infinite
Quantum Systems}, \\ Springer, 1993.

\bibitem{Dev}  Devaney, R.L.,  \emph{An Introduction to Chaotic dynamical
Systems}, \\ Benjamin, 1986.

\bibitem{Has}  Hasegawa, H., Dynamical formulation of quantum level
ststistics, Open Systems and Information dynamics, {\bf 4}, 350, 1997.

\bibitem{IKO}  Ingarden, R.S., Kossakowski, A. and Ohya, M.,
\emph{Information Dynamics and Open Systems}, Kluwer Academic
Publishers, 1997.

\bibitem{KO}  Kosaka, M., and Ohya, M., A study of chaotic phenomena by
information dynamics (in Japanese), IEICE, {\bf J80-A}, No.12, 2138, 1997.

\bibitem{Mis}  Misiurewicz, M., Absolutely continuous measutres for
certain maps of interval, Publ. Math. IHES, Vol.53, 17, 1981.

\bibitem{O1}  Ohya, M., Information dynamics and its applications to
optical communication processes, Lecture Note in Physics, {\bf 378},
81, 1991.

\bibitem{O2}  Ohya, M., Complexity and fractal dimensions for quantum
states, Open Systems and Information Dynamics, {\bf 4}, 141, 1997.

\bibitem{O3}  Ohya, M., Complexities and their applications to
characterization of chaos, International Journal of Theoretical Physics,
{\bf 37}, No.1, 495, 1998.

\bibitem{Tod} Toda, M., Crisis in chaotic scattering of a highly excited
van der Waals complex, Physical Review Letters, {\bf 74}, No.14,
2970, 1995.
\end{thebibliography}
\end{document}